\documentclass[
reprint,
superscriptaddress,
amsmath,amssymb,
aps,
twocolumn
]{revtex4-2}

\usepackage{graphicx}
\usepackage{dcolumn}
\usepackage{bm}
\usepackage[bookmarks=false,
 breaklinks=true,pdfborder={0 0 0},pdfborderstyle={},backref=false,colorlinks=true]
 {hyperref}
\usepackage{mathtools}
\usepackage{physics}
\usepackage{amsmath}
\usepackage{bm}
\usepackage{CJKutf8}
\usepackage{subcaption}
\usepackage[justification=raggedright, singlelinecheck=false]{caption}
\usepackage{xcolor}
\usepackage{booktabs}
\usepackage{multirow}
\usepackage{array}
\usepackage{hhline}

\definecolor{Tomato}{RGB}{255,99,71}
\newcommand{\dash}{\ensuremath{-}}

\begin{document}

\title{Classical analog of the T. D. Lee model for renormalization}

\author{Kang Xu (\begin{CJK}{UTF8}{gbsn}许康\end{CJK})}
\affiliation{Graduate School of China Academy of Engineering Physics, Beijing 100193,
China}

\author{C. P. Sun  (\begin{CJK}{UTF8}{gbsn}孙昌璞\end{CJK})}
\email{suncp@gscaep.ac.cn}

\affiliation{Graduate School of China Academy of Engineering Physics, Beijing 100193,
China}

\date{\today}

\begin{abstract}
While divergence and renormalization of physical quantities are frequently encountered in quantum field theory (QFT), they are not necessarily quantum-specific characteristics. We show in this paper that there exists a classical counterpart of the Lee model which is the model of coupled harmonic oscillators (CHO). It is demonstrated that the frequency divergence in this classical model precisely replicates the phenomenon of mass divergence in the Lee model, as does the corresponding renormalization procedure. Considering the arbitrariness in renormalization schemes, we establish necessary conditions that a general renormalization must satisfy for the model of two coupled oscillators which corresponds to the single-mode Lee model. Furthermore, we analyze the classical analog of $N\text{-}\theta$ scattering process and show that the dependence of scattering strength on the cutoff mode mirrors that of the quantum case. These findings challenge the quantum-centric view of mass renormalization in the Lee model and offer new insights into the classical–quantum correspondence in renormalization theories.
\end{abstract}

\maketitle

\section{\label{sec:intro}Introduction}
The Lee Model \cite{lee1954some}, proposed in 1954, describes interactions between two species of fermions denoted as $V,N$ particles, and a bosonic field $\theta$. The Lee model provided the first analytical demonstration of the essential requirement of renormalization in quantum field theory (QFT) even though the model is exactly solvable. This finding reveals that the divergence problem originates from the model's intrinsic structure, rather than solely from the perturbation theory. Furthermore, the Lee model also inspired systematic investigations on the properties of the bound states, together with their essential roles in renormalization procedures\cite{kallen1955mathematical}.

Beyond its fundamental contributions to QFT, the Lee model profoundly impacts many realms in physics. It was first seen in the Lee model that for a fixed renormalized coupling the bare coupling becomes imaginary beyond a certain ultraviolet cut-off~\cite{brodsky1998quantum}. This behavior is accompanied by the emergence of a ghost state, defined as a negative-norm bound state in the Lee model, which has attracted considerable attention in the study of quantum theories with indefinite metric \cite{kallen1955mathematical,sudarshan1961quantum,lee1969negative,nakanishi1972indefinite}. The Lee model provides a framework to demonstrate that, non-Hermitian Hamiltonians with $\mathcal{PT}$ symmetry can still consistently formulate QFT through metric redefinition and Hilbert space reconstruction \cite{bender2005ghost,bender2007making,shi2009recoveringunitarityleemodel,mintz2025oscillators,felski2021towards,croney2023renormalization,mavromatos2024chern,teufel2021hamiltonians}. The resulting theory restores the unitarity of $S$-matrix and positive-definite norms of quantum states. In studies of unstable systems, the Lee model is often employed to investigate the dynamics of the decay process \cite{glaser1956model,araki1957quantum,chiu1977time,horwitz1995unstable,nakazato1996temporal,giacosa2012non,giacosa2022multichannel}. In the fields of atomic physics and condensed matter physics, the Lee model is frequently denoted as the Friedrichs-Lee model \cite{friedrichs1948perturbation,facchi2021spectral,matsyshyn2023fermi}. Notably, under certain parameter regimes, the Lee model exhibits identical physical characteristics to several prominent  physical models, such as the Tavis-Cummings (TC), Fano, and Anderson models \cite{zhou2006electromagnetic,shi2009lehmann,shi2009twophotonscatteringdimensionlocalized,xu2010photonic}. 
For example, the Lee model converts to the TC model if we map the two fermions to two-level states and interpret the boson as photon in TC model. Due to its structural simplicity, the Lee model is particularly suitable for pedagogical exposition of fundamental concepts in QFT , including calculations of $S$-matrix and renormalization procedures \cite{barton1963introduction,schweber2005introduction}.

The properties of the Lee model, especially the renormalization of divergent observables, are often studied within QFT and many-body physics. This seems to be well understood, but we find that such divergence problems persist similarly even in classical mechanics, where some kind of renormalization remains necessary. In this paper, we consider a classical counterpart of the Lee model in the so-called $N\theta/V$ sector by mapping it to a model of coupled harmonic oscillators (CHO) through artificial coupling. Through this correspondence, we investigate the classical counterparts of the mass renormalization and the suppression of scattering amplitudes of $N\theta$ scattering, respectively.

Firstly, we demonstrate that the problem of mass divergence in the Lee model also emerges in the corresponding CHO model , where the divergent quantity is the lowest natural frequency of this system. It is found that the divergent natural frequency is negative, implying an unphysical Hamiltonian. As a demonstration of this non-physicality, let us examine a two-coupled harmonic oscillator (TCHO) model characterized by coordinates $x$ and $y$, unit masses, and respective frequencies $\omega_x$ and $\omega_y$. If the interaction between the two oscillators is simply modeled as $\lambda xy$ in Hamiltonian, the potential landscape will transform into an inverted harmonic oscillator configuration in strong coupling regime ($\lambda^2 \geq \omega_x\omega_y$). Such potential will induce unbounded particle acceleration, which implies that the Hamiltonian is unphysical. To restore physical plausibility of the Hamiltonian in strong coupling regime, renormalization terms like $\Delta\omega^2x^2$ or $\Delta\omega^2y^2$ are required. Similarly, frequency renormalization is necessary in the aforementioned CHO model, which proved to be identical to the mass renormalization in the Lee model.

Secondly, it is recognized that choices of renormalization schemes possess a certain arbitrariness. The same applies to the Lee model. For analytical simplicity, we investigate TCHO systems with artificial $xy$-$p_xp_y$ coupling which corresponds to the single-mode Lee model. The general renormalization schemes for such systems is examined. By requiring the positive-definiteness of the renormalized Hamiltonian, we establish necessary conditions that the general renormalization schemes must satisfy.

We also establish a classical analogue of the $N\text{-}\theta$ scattering process using our model by regarding the oscillator amplitudes as an indication of occupation probability of scattered states. The analogous scattering strength shows similar cut-off momentum dependence to the actual $N\text{-}\theta$ scattering cross-section. These results demonstrate that mass renormalization and the suppression of the scattering amplitudes in the Lee model are not exclusively quantum features but also manifest in a classical context, revealing their origins in the model's structural configuration of classical nature.

The paper is organized as follows. In Sec.~\ref{sec:II}, we establish the classical correspondence of the Lee model. In Sec.~\ref{sec:III}, we demonstrate the classical perspective of mass renormalization. In Sec.~\ref{sec:IV}, we discuss a general renormalization scheme for two coupled harmonic oscillators, highlighting that the single-mode Lee model serves as a specific example of this broader framework. In Sec.~\ref{Sec: V} we demonstrate the analog of $N\text{-}\theta$ scattering in our classical model which shows a similar "scattering" behavior of cut-off momentum dependence to the $N\text{-}\theta$ scattering cross section. We finish in Sec.~\ref{sec conclusion} with the conclusion and outlooks.

\section{\label{sec:II}Classical correspondence of Lee model}
In this section, we will introduce the classical correspondence of the Lee model by mapping it to a CHO model. The Lee model describes a conversion process among three kinds of particles, $V, N$ and $\theta$, represented as
\begin{equation*}
    V\rightleftharpoons N+\theta_{\bm{k}},
\end{equation*}
where $V$ and $N$ are fermions and $\theta_{\bm{k}}$ are bosons with 3-dimensional momentum $\bm{k}$. Such process is governed by Hamiltonian
\begin{eqnarray}\label{Lee H}
    H_{\text{Lee}}&=&m_V V^\dagger V+m_N N^\dagger N+\sum_{\bm{k}}\omega_{\bm{k}} a_{\bm{k}}^\dagger a_{\bm{k}}\nonumber\\
    &&+\sum_{\bm{k}}\frac{g}{\sqrt{2\omega_{\bm{k}}\Omega}}(VN^\dagger a_{\bm{k}}^\dagger+h.c.)~.
\end{eqnarray}
Here $g$ is the coupling constant, $\Omega$ the space volume. $V^\dagger$, $N^\dagger$ and $a_{\bm{k}}^\dagger$ are the creation operators of $V$, $N$, and $\theta_{\bm{k}}$ particles, respectively, obeying the commutation (anti-commutation) rules for bosons (fermions), $\left[a_{\bm{k}},a_{\bm{k}'}\right]=\delta_{\bm{k},\bm{k}'},~ \{V,V^\dagger\}=1,~\{N,N^\dagger\}=1$~. The dispersion relation of $\theta_{\bm{k}}$ is $\omega_{\bm{k}}^2=\mu^2+\bm{k}^2$, where $\mu$ is the rest mass of $\theta$ particle. The rest masses of the $V$ and $N$ particles ($m_V$ and $m_N$) are sufficiently large that the recoil from $\theta$-particle scattering can be neglected, i.~e.~the $V$ and $N$ particles can be treated as stationary. Therefore the momentum dependence of energy of the $V,N$ particles can be neglected. Furthermore, the mass $m_V$ is set to be less than $m_N+\mu$ (i.e., $m_V<m_N+\mu$) to ensure the $V$-particle does not spontaneously decay into $N$ and $\theta$. This stability is essential for the role of bound states in renormalization procedures \cite{kallen1955mathematical}.

There are two  conserved quantities
\begin{eqnarray*}
    n_1&=&V^\dagger V+N^\dagger N\nonumber~,\\
    n_2&=&N^\dagger N-\sum_{\bm{k}} a_{\bm{k}}^\dagger a_{\bm{k}}~.
\end{eqnarray*}
It can be verified that $[H,n_1]=[H,n_2]=0$. Consequently, the Hilbert space can be decomposed into eigen-spaces defined by different quantum number $n_1,n_2$. In each subspace (often referred to as a sector in the Lee model), the Hamiltonian Eq.~\eqref{Lee H}. is analytically soluble.

Among all these subspaces, the sector with $n_1=1$, $n_2=0$ (denoted as $N\theta/V$-sector), has been studied in detail for it is the simplest invariant subspace to handle while preserving key QFT features, such as mass and coupling constant renormalization. The $N\theta/V$ sector is spanned by $\{\ket{V},\ket{N\theta_{\bm{k}_1}},\cdots,\ket{N\theta_{\bm{k}_n}}\}$, where $\ket{V}=V^\dagger\ket{0}$ and $\ket{N\theta_{\bm{k}}}=N^\dagger a_{\bm{k}}^\dagger\ket{0}$ , $\ket{0}$ represents the vacuum state. In this physical basis, the matrix form of Hamiltonian~\eqref{Lee H} reads as:
  \begin{equation}\label{MatrixH}
    H_\text{Lee}=
    \begin{pmatrix}
        m_V & \frac{g}{\sqrt{2\omega_{{\bm{k}}_1}\Omega}} & \cdots & \frac{g}{\sqrt{2\omega_{{\bm{k}}_n}\Omega}} \\
        \frac{g}{\sqrt{2\omega_{{\bm{k}}_1}\Omega}} & m_N+\omega_{{\bm{k}}_1} & \bm{0} & 0 \\
        \vdots & \bm{0} & \ddots & \bm{0} \\
        \frac{g}{\sqrt{2\omega_{{\bm{k}}_n}\Omega}} & 0 & \bm{0} & m_N+\omega_{{\bm{k}}_n}
    \end{pmatrix}~,
    \end{equation}
which contains non-zero elements only in the first row, first column, and along the diagonal.

With such representation, the Lee model can be mapped to the single excitation subspace of a bosonic model whose Hamiltonian is
\begin{eqnarray}\label{CHO H}
    H&=&m_V b^\dagger b+\sum_{\bm{k}}(\omega_{\bm{k}}+m_N)c_{\bm{k}}^\dagger c_{\bm{k}}\nonumber\\
    &&+\sum_{\bm{k}}\frac{g}{\sqrt{2\omega_{\bm{k}}\Omega}}\left(bc_{\bm{k}}^\dagger+b^\dagger c_{\bm{k}}\right),
\end{eqnarray}
where $b~,~b^\dagger~,~c_{\bm{k}}~,~c_{\bm{k}}^\dagger$~are bosonic operators obeying $ \left[b,b^\dagger\right]=1~,\left[c_{\bm{k}},c_{\bm{k}'}^\dagger\right]=\delta_{\bm{k},\bm{k'}}~.$ The total particle number operator $\hat{n}=b^\dagger b+\sum_{\bm{k}} c_{\bm{k}}^\dagger c_{\bm{k}}$ commutes with the Hamiltonian, thus the total particle number is conserved in this model. In the single excitation subspace, (i.~e.~ the subspace where the eigenvalue of $\hat{n}$ is 1), via the following basis mapping
\begin{equation*}
    b^\dagger\ket{0}\to \ket{V},\quad c_{\bm{k}}^\dagger\ket{0}\to \ket{N\theta_{\bm{k}}},
\end{equation*}
we find the matrix representation of $H$ identical to that of $H_\text{Lee}$ in its $N\theta/V$-sector (Eq.\eqref{MatrixH}). This means that the simplified Hamiltonian Eq.~\eqref{CHO H} captures the physics of the Lee model within the $N\theta/V$-sector.

\begin{figure}
\centering
\includegraphics[width=0.5\columnwidth]{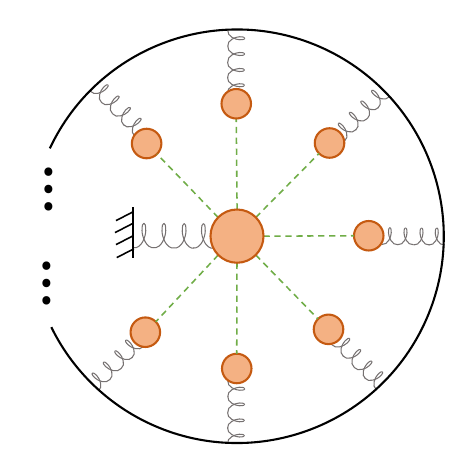}
\caption{\label{Fig classical model}The classical interpretation of the Lee model within $N\theta/V$ sector: A classical model described by Hamiltonian Eq.~\eqref{config rep} consists of a central harmonic oscillator with frequency $m_V$ coupled to a set of environment oscillators with frequency spectrum $m_N+\omega_{\bm{k}}$~, the dashed lines represent the $xx_{\bm{k}}\text{-}pp_{\bm{k}}$ coupling between the central oscillator and environment oscillators.}
\end{figure}

Through this correspondence, we can establish a classical counterpart of the Lee model. In fact, we can rewrite the Hamiltonian~\eqref{CHO H} in configuration space by
\begin{eqnarray}\label{b to x}
    b&=&\sqrt{\frac{1}{2}}(x+ip)~,\\
    c_{\bm{k}}&=&\sqrt{\frac{1}{2}}(x_{\bm{k}}+ip_{\bm{k}})~,
\end{eqnarray}
where $x,~p,~x_{\bm{k}}$ and $~p_{\bm{k}}$ denote position and momentum operators obeying $[x,p]=i~,[x_{\bm{k}},p_{\bm{k}'}]=i\delta_{\bm{k},\bm{k}'}$~(with $\hbar=1$). Omitting constant energy terms, the Hamiltonian~\eqref{CHO H} becomes:
\begin{align}\label{config rep}
         H=&\frac{1}{2}m_V(p^2+x^2)+\sum_{\bm{k}}\frac{1}{2}(m_N+\omega_{\bm{k}})(p_{\bm{k}}^2+x_{\bm{k}}^2)\nonumber\\
         &+\sum_{\bm{k}}\frac{g}{\sqrt{2\omega_{\bm{k}}\Omega}}\left(xx_k+pp_{\bm{k}}\right)~.
    \end{align}
This Hamiltonian describes a system comprising a central harmonic oscillator $(x,p)$ with frequency $m_V$ and environmental oscillators $(x_{\bm{k}},p_{\bm{k}})$ with frequencies $m_N+\omega_{\bm{k}}$. The central oscillator couples to each environmental mode via position-momentum interactions, as illustrated in Fig.~\ref{Fig classical model}~. Thus, the classical counterpart of the Lee model within the $N\theta/V$-sector is identified with the model of coupled harmonic oscillators.

\section{\label{sec:III} A Classical Perspective of Mass Renormalization}
In last section we have shown that the Lee model can be mapped to a CHO model in a special subspace. In this section, we examine this classical model to investigate one of the critical issues in the Lee model, which is mass renormalization. A prominent divergent observable in the Lee model is the bound-state energy $m_V' $ which is usually defined as the physical mass of the $V$ particle. This divergence arises from unregulated ultraviolet modes of the bosonic field $\theta_{\bm{k}}$ coupling to the $N$ and $V$ particles. The introduction of mass renormalization serves to eliminate this divergence and consequently yields a finite physical mass $m_V'$~. In the subsequent studies of the Lee model, Källén et~al.~\cite{kallen1955mathematical,glaser1956model,weinberg1956n} imposed an ultraviolet cutoff $k_c$ on the $\bm{k}$-modes (where $|\bm{k}| \leq k_c$) to avoid this divergence.  However, mass renormalization is still necessary to make $m_V'$ cut-off independent. By comparing the divergences in the Lee model with those in its classical counterpart, we investigate whether the mass divergence in the Lee model originates from quantum mechanical effects.

In the Lee model, the physical mass of the $V$ particle $m_V'$ is the lowest eigen-energy within the $N\theta/V$-sector, which reduces to the original mass $m_V$ when the coupling constant approaches zero, i.~e.~$g\to0$. Correspondingly, in the CHO model we examine the divergence of the lowest eigen-frequency of the system. The matrix representation of Hamiltonian~\eqref{config rep} reads as
\begin{equation}\label{config matrix}
H=\left(\bm{p}^T,  \bm{x}^T\right)
    \begin{pmatrix}
        H_\text{block} & \bm{0}\\
        \bm{0} & H_\text{block}
    \end{pmatrix}
    \begin{pmatrix}
        \bm{p}\\
        \bm{x}
    \end{pmatrix},
\end{equation}
where $\bm{p}=(p,p_{{\bm{k}}_1},\ldots,p_{{\bm{k}}_n})^T$, $\bm{x}=(x,x_{{\bm{k}}_1},\ldots,x_{{\bm{k}}_n})^T$, and
\begin{equation}
H_\text{block}=
         \begin{pmatrix}
        m_V & \frac{g}{\sqrt{2\omega_{{\bm{k}}_1}\Omega}} & \cdots & \frac{g}{\sqrt{2\omega_{{\bm{k}}_n}\Omega}} \\
        \frac{g}{\sqrt{2\omega_{{\bm{k}}_1}\Omega}} & m_N+\omega_{{\bm{k}}_1} & \bm{0} & 0 \\
        \vdots & \bm{0} & \ddots & \bm{0} \\
        \frac{g}{\sqrt{2\omega_{{\bm{k}}_n}\Omega}} & 0 & \bm{0} & m_N+\omega_{{\bm{k}}_n}
    \end{pmatrix}~.
    \end{equation}
Let $U$ be an $(n+1)\times(n+1)$ orthogonal matrix, i.~e.~$U^TU = I$. It generates a canonical transformation of the conjugate variables $(\bm{p}^T,\bm{x}^T)$ as follows
\begin{equation}
    \begin{pmatrix}
        \bm{P}\\
        \bm{X}
    \end{pmatrix}
    =
    \begin{pmatrix}
        U & \bm{0}\\
        \bm{0} & U
    \end{pmatrix}
    \begin{pmatrix}
        \bm{p}\\
        \bm{x}
    \end{pmatrix}~,
\end{equation}
where $\bm{P}=(P,P_{{\bm{k}}_1},\ldots,P_{{\bm{k}}_n})^T$ and $\bm{X}=(X,X_{{\bm{k}}_1},\ldots,X_{{\bm{k}}_n})^T$. The Hamiltonian matrix in the new basis $(P^T,X^T)$ is given by
\begin{equation}
H=
    \begin{pmatrix}
        UH_\text{block}U^{-1} & \bm{0}\\
        \bm{0} & UH_\text{block}U^{-1}
    \end{pmatrix}.
\end{equation}
Diagonalizing $H_\text{block}$ via $U$ yields the natural frequency $\omega$ determined by the secular equation $\det(\omega I - H_\text{block}) = 0$, namely
\begin{equation}
        \begin{vmatrix}
     m_V-\omega & g(2\omega_{{\bm{k}}_1}\Omega)^{-1/2} & \cdots & g(2\omega_{{\bm{k}}_n}\Omega)^{-1/2}\\
        g(2\omega_{{\bm{k}}_1}\Omega)^{-1/2} & m_N+\omega_{{\bm{k}}_1}-\omega & &\\
        \vdots & & \ddots &\\
        g(2\omega_{{\bm{k}}_n}\Omega)^{-1/2} & & & m_N+\omega_{{\bm{k}}_n}-\omega 
\end{vmatrix}=0~,
 \end{equation}
which can be simplified to
\begin{equation}\label{eigenvalue eq}
  h(\omega) \equiv \omega - m_V - \sum_{\bm{k}} \frac{g^2}{2\Omega\omega_{\bm{k}}(\omega - m_N - \omega_{\bm{k}})} = 0~.
\end{equation}
This is just identical to the eigenvalue equation of the Lee model Hamiltonian (see Appendix.~\ref{app:mass renormalization} for more details), which signifies that the eigen-frequencies of the CHO model exhibit a one-to-one correspondence to the eigen-energies of the Lee model. Consequently, the lowest natural frequency of the CHO model has the same value with the physical mass of the $V$ particle, which is $m_V'$~.

\begin{figure}
\centering
\includegraphics[width=\columnwidth]{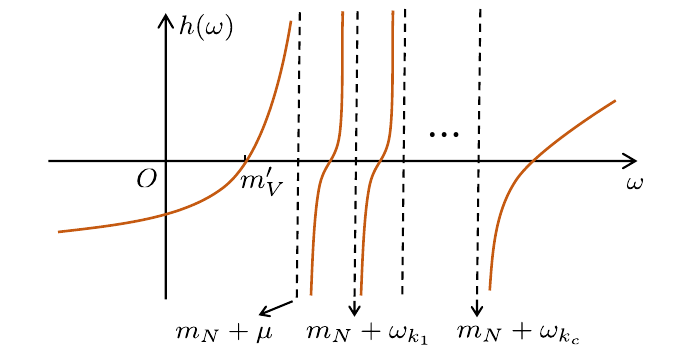}
\caption{\label{Fig secular function}The graph of secular function $h(\omega)$. The zero points of the function represent the eigen-frequencies of the system described by Hamiltonian~\eqref{config rep}.}
\end{figure}

The secular function \( h(\omega) \) are singular at \( \omega = m_N + \omega_{\bm{k}} \) for each \( \bm{k} \), among which the lowest one is at \( \omega = m_N + \mu \) (see Fig.~\ref{Fig secular function} for the graph of function $h(\omega)$). Because $h'(\omega)>0$, it can be verified that there is one and only one solution of Eq.~\eqref{eigenvalue eq} in region $\omega<m_N+\mu$, which is $m_V'$.

The summation in Eq.~\eqref{eigenvalue eq} diverges because the magnitudes of the three-dimensional vectors $\bm{k}$ are unbounded. Consequently, the natural frequency $m_V' \to -\infty$. This result indicates that the ultraviolet divergence persists in the classical model; the mass divergence in the Lee model is not a quantum mechanical effect but rather a classical one.

The divergence $m_V'$ means that the fundamental frequency of the system's normal modes are infinitely high. Besides, the negative value of the natural frequency $m_V'$ will leads to an unstable system described by Hamiltonian
\begin{equation}\label{H'}
    H'=-\frac{1}{2}|m_V'|(P^2 + X^2) + \sum_{\bm{k}} \frac{1}{2}(m_N + \omega_{\bm{k}}')(P_{\bm{k}}^2 + X_{\bm{k}}^2)~.
\end{equation}
This Hamiltonian is unphysical since its energy is unbounded from below. From a quantum mechanical perspective, a negative natural frequency corresponds to excitations with negative energy, consequently the system will inevitably develop infinite excitations. Evidently, these unphysical outcomes cannot represent actual physics, indicating that the form of coupling in the model is an approximation that works in weak coupling regime but fails in strong coupling regime. Renormalization modifies the Hamiltonian by introducing counter terms to cancel out the divergence of $m_V'$ and restore its positive value.

More specifically, we can introduce a term $\Delta H$ into the original Hamiltonian~\eqref{config rep},
\begin{equation}\label{D H}
    \Delta H = \frac{1}{2}\delta m(p^2 + x^2).
\end{equation}
The renormalized Hamiltonian $H_R = H + \Delta H$ then reads
\begin{align}
    H_R =&\frac{1}{2}(m_V + \delta m)(p^2 + x^2)+\sum_{\bm{k}}\frac{1}{2}(m_N + \omega_{\bm{k}})(p_{\bm{k}}^2 + x_{\bm{k}}^2)\nonumber\\
    & + \sum_k\frac{g}{\sqrt{2\omega_{\bm{k}}\Omega}}(xx_{\bm{k}} + pp_{\bm{k}})~.
\end{align}
The corresponding eigenvalue equation now becomes
\begin{equation}\label{modified eigen fun}
    \omega - m_V - \delta m - \sum_{\bm{k}} \frac{g^2}{2\Omega \omega_{\bm{k}} (\omega - m_N - \omega_{\bm{k}})} = 0~.
\end{equation}
By setting the parameter $m_V$ to coincide with the physical frequency  (i.~e.~requiring $ m_V$ to be a solution of Eq.~\eqref{modified eigen fun}) , we have the renormalization coefficient:
\begin{equation}\label{d m}
    \delta m = \sum_{\bm{k}} \frac{g^2}{2\Omega \omega_k (m_N + \omega_k - m_V)}~.
\end{equation}
This coefficient precisely matches that of mass renormalization in the Lee model. Furthermore, in the second quantization formalism, the renormalization term~\eqref{D H} introduced in our classical model reads as
\begin{equation}
    \Delta H = \delta m b^\dagger b~.
\end{equation}
This shares the same form as the mass renormalization term $\Delta H_\text{Lee} = \delta m V^\dagger V$ in the Lee model. These results demonstrate that both the mass divergence and its renormalization arise from the classical nature of the model rather than a quantum one.

\section{\label{sec:IV}General Renormalization for Coupled Oscillator Systems}
In the classical perspective of renormalization discussed above, specific renormalization terms are introduced to eliminate the model's energy divergence. However, the choices of these terms are not unique. Besides Eq.~\eqref{D H}, there are many alternative renormalization schemes that can also resolve the mass divergence. For instance, the coefficients of $p^2$ and $x^2$ in Eq.~\eqref{D H} are not necessarily identical, or the objects to be renormalized may be chosen as the environmental oscillators instead of the central oscillator. The specific choice of renormalization~\eqref{D H} is made solely for the sake of computational simplicity. Notably, not all added terms result in a physical Hamiltonian which describes a stable system. In this section we derive the necessary constraints for general renormalization schemes of CHO systems with $xx$-$pp$ coupling to guarantee that  the renormalized Hamiltonian is physically meaningful, that is, the Hamiltonian has a finite lower bound of energy.

For simplicity, we consider a TCHO model with $xx$-$pp$ coupling, whose Hamiltonian can be split into two parts, $H_2 = H_p + H_x$,
\begin{equation}\label{TCHO H}
    \begin{aligned}
    H_p &= \frac{1}{2}\omega_1p_1^2 + \frac{1}{2}\omega_2p_2^2 + g_p p_1p_2~, \\
    H_x &= \frac{1}{2}\omega_1x_1^2 + \frac{1}{2}\omega_2x_2^2 + g_x x_1x_2~.
    \end{aligned}
\end{equation}

Here $x_1,x_2$ and $p_1,p_2$ are the positions and momenta of the two oscillators respectively. The Lee model with a monochromatic bosonic field $\theta$ (termed single mode Lee model) serves as a special case of TCHO, where $g_x=g_p$. Both $H_p$ and $H_x$ become non-positive-definite when $g_x^2, g_p^2 > \omega_1\omega_2$, that is, there is no lower bound on the total energy which means the system is unstable.  To restore the stability of this system, or mathematically speaking, to ensure the Hamiltonian's positive-definiteness, we renormalize the Hamiltonian as $H_{2R} = H_p + \Delta H_p + H_x + \Delta H_x$~. The renormalization terms can be generally written as
\begin{align}
    \Delta H_p &= \frac{1}{2}a_1 p_1^2 + \frac{1}{2}a_2 p_2^2~, \\
    \Delta H_x &= \frac{1}{2}b_1 x_1^2 + \frac{1}{2}b_2 x_2^2~,
\end{align}
where coefficients $a_i$ and $b_i$ are to be determined.

\begin{figure}
\includegraphics[width=\columnwidth]{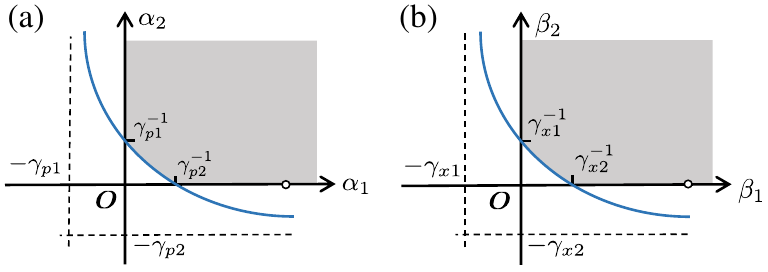}
\caption{\label{allowed area} A renormalization scheme comprises four parameters $\alpha_1$, $\alpha_2$, $\beta_1$, $\beta_2$ defined in Eq.~\eqref{dimensionless coeff}. (a) The parameter space of $\alpha_1$-$\alpha_2$. The shaded region represents the range of parameters satisfying condition~\eqref{dim cond 1}. (b) The parameter space of $\beta_1$-$\beta_2$. The shaded region represents the range of parameters satisfying condition~\eqref{dim cond 2}.
The parameters of an admissible renormalization scheme must lie within the shaded regions of both of the diagrams (a) and (b) to ensure the positive-definiteness of the renormalized Hamiltonian. 
In the case of the classical counterpart of the singel-mode Lee model, the parameters take the values in Eq.~\eqref{single model Lee para}, which are marked as the hollow points in the figure.}
\end{figure}

The positive-definiteness for $H_{2R}$ requires the positive-definiteness of its Hessian matrix, which means
\begin{equation}\label{p-d condition 1}
\begin{cases} 
(\omega_1+a_1)(\omega_2+a_2) - g_p^2 > 0~, \\ 
\omega_1+a_1 > 0~, \\
\omega_2+a_2 > 0~, 
\end{cases}
\end{equation}
and
\begin{equation}\label{p-d condition 2}
\begin{cases} 
(\omega_1+b_1)(\omega_2+b_2) - g_x^2 > 0~, \\ 
\omega_1+b_1 > 0~, \\
\omega_2+b_2 > 0~.
\end{cases}
\end{equation}
In the strong coupling regime where $g_x^2,g_p^2 \gg \omega_1,\omega_2$, these conditions reduce to
\begin{equation}\label{dim cond 1}
\begin{cases} 
\gamma_{p2}\alpha_1 + \gamma_{p1}\alpha_2 + \alpha_1\alpha_2 \geq 1~, \\ 
\alpha_1,\alpha_2 \geq 0~,
\end{cases}
\end{equation}
and
\begin{equation}\label{dim cond 2}
\begin{cases} 
\gamma_{x2}\beta_1 + \gamma_{x1}\beta_2 + \beta_1\beta_2 \geq 1~, \\ 
\beta_1,\beta_2 \geq 0~,
\end{cases}
\end{equation}
where we have defined dimensionless parameters scaled by coupling constants,
\begin{align}\label{dimensionless coeff}
&\begin{cases} 
\gamma_{p1} \equiv \dfrac{\omega_1}{|g_p|},\\ 
\gamma_{p2} \equiv \dfrac{\omega_2}{|g_p|},
\end{cases}
&&\begin{cases} 
\gamma_{x1} \equiv \dfrac{\omega_1}{|g_x|},\\ 
\gamma_{x2} \equiv \dfrac{\omega_2}{|g_x|},
\end{cases}\\
&\begin{cases} 
\alpha_1 \equiv \dfrac{a_1}{|g_p|},\\ 
\alpha_2 \equiv \dfrac{a_2}{|g_p|},
\end{cases}
&&\begin{cases} 
\beta_1 \equiv \dfrac{b_1}{|g_x|},\\ 
\beta_2 \equiv \dfrac{b_2}{|g_x|}.
\end{cases}
\end{align}
The diagrams of the parameter space of $\alpha_i$ and $\beta_i$ are illustrated in Fig~\ref{allowed area}, where the shaded regions represent the parameter ranges satisfying the conditions~\eqref{dim cond 1} and \eqref{dim cond 2} respectively. To achieve a positive-definite Hamiltonian, the parameters $(\alpha_1,\alpha_2)$ and $(\beta_1,\beta_2)$ of a renormalization scheme must simultaneously lie within the shaded regions.

The classical counterpart of the single-mode Lee model provides a specific example of the above discussion. The single-mode Lee model Hamiltonian is:
\begin{eqnarray*}
    H_{\text{Lee}}' &=& m_V V^\dagger V + m_N N^\dagger N + \omega_k a_k^\dagger a_k \\
    && + \frac{g'}{\sqrt{2\omega_k\Omega}}(VN^\dagger a_k^\dagger + \text{h.c.})~.
\end{eqnarray*}
As shown in Section~\ref{sec:II}, this maps to the coupled oscillator model with Hamiltonian
\begin{equation}
    H_2' = \frac{1}{2}\omega_1(p_1^2 + x_1^2) + \frac{1}{2}\omega_2(p_2^2 + x_2^2) + g(x_1x_2 + p_1p_2)~,
\end{equation}
where $g = g'/\sqrt{2\omega_k\Omega} \omega_1=m_V$, and $ \omega_2=m_N+\omega_k$. According to Eq.~\eqref{D H} and \eqref{d m}, the renormalization term of $H_2'$ is given by
\begin{equation}\label{Delta H in single Lee}
    \Delta H = \frac{1}{2}\left(\frac{g^2}{\omega_2 - \omega_1}\right)(x_1^2 + p_1^2)~,
\end{equation}
The dimensionless coefficients in this renormalization term are
\begin{equation}\label{single model Lee para}
    (\alpha_1,\alpha_2,\beta_1,\beta_2) = ((\gamma_2-\gamma_1)^{-1},0,(\gamma_2-\gamma_1)^{-1},0),
\end{equation}
which lie exactly on the boundaries of the admissible parameter space, represented by the hollow points in Fig.~\ref{allowed area}.

As discussed earlier, there are alternative renormalization schemes besides the one adopted in the Lee model. For example, we can consider the following renormalization terms instead of Eq.\eqref{Delta H in single Lee}:
\begin{equation}
\begin{aligned} 
    \Delta H_x &= \frac{1}{2}|g_p|(x_1^2 + x_2^2)~, \\
    \Delta H_p &= \frac{1}{2}|g_x|(p_1^2 + p_2^2)~.
\end{aligned}
\end{equation}
The corresponding dimensionless parameters $(\alpha_1,\alpha_2,\beta_1,\beta_2) = (1,1,1,1)$ meet the conditions in Eq. \eqref{dim cond 1} and\eqref{dim cond 2}. The resulting renormalized Hamiltonian
\begin{equation}
\begin{aligned}
    H_{2R} =& \frac{1}{2}\omega_1(p_1^2 + x_1^2) + \frac{1}{2}\omega_2(p_2^2 + x_2^2) \\
    &+ \frac{1}{2}|g_p|(p_1 - p_2)^2 + \frac{1}{2}|g_x|(x_1 - x_2)^2.
\end{aligned}
\end{equation}
This Hamiltonian is positive-definite since all terms are squared quantities, ensuring the stability of the system.

\section{\label{Sec: V}Classical analog of $\bm{N\text{-}\theta}$ scattering}
Besides mass renormalization, another significant issue in the Lee model is the vanishing of the cross section. As cutoff momentum $k_c$ tends to infinity, the cross section of $N+\theta_k\rightarrow N+\theta_{k'}$ scattering is suppressed to zero (See Appendix.~\ref{app:scattering} for details). In previous sections we have demonstrated that the mass divergence is a classical property of the model, this raises the question of whether the suppression of cross section also corresponds to a classical phenomenon. Motivated by this association, we investigate the mode cutoff dependence of "scattering" in our classical model, which parallels $N\theta$ scattering in the Lee model.

\begin{figure}
\includegraphics[width=\columnwidth]{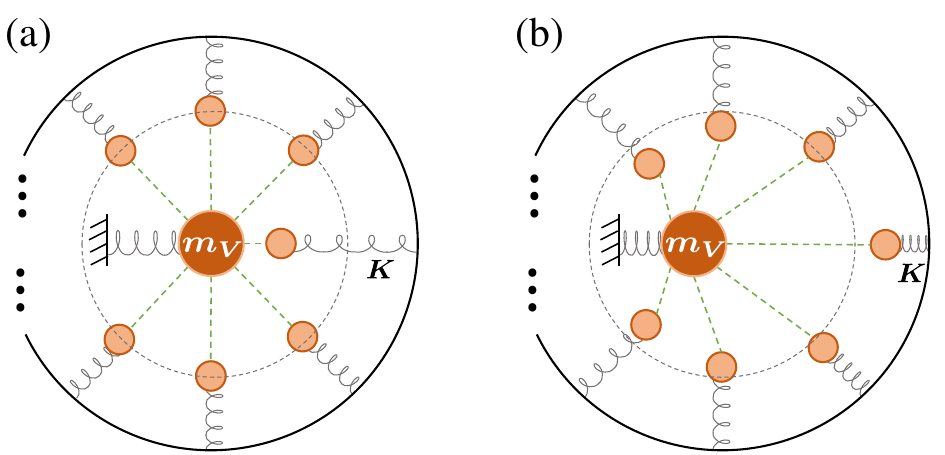}
\caption{\label{Fig scatter}The classical analog of $N\text{-}\theta$ scattering in the Lee model: (a)Initial configuration for  the environmental oscillator with momentum parameter $\bm{K}$ is displaced from equilibrium, while others remain at rest at their equilibrium positions indicated by the dashed circle. (b) After a period of temporal evolution governed by Hamiltonian~\eqref{config rep}, oscillators in other modes such as the $\bm{k'}$ mode are excited owing to the driving influence of the oscillator initially prepared in mode $K$.}
\end{figure}

Our classical analogy is formulated as follows. In the Lee model, suppose the in-state is $\ket{N\theta_{\bm{K}}}$, i.~e.~the incoming particles consists of an $N$ particle and a $\theta$ particle with momentum $K$. Correspondingly, in our classical set up, only the $\bm{K}$-mode environmental oscillator is displaced from equilibrium position, while others remain stationary (see Fig.~\ref{Fig scatter}).
Subsequent distribution of oscillator amplitudes at a later time $t$ reflect the probability distribution of final scattering states: larger amplitudes indicate stronger "transitions" to corresponding momentum modes.

Building upon this correspondence, we now derive the dynamics of environmental oscillators. The equations of motion for the oscillators governed by  Hamiltonian~\eqref{config rep} are
\begin{widetext}
    \begin{align}
    &\ddot{x}=-(m_V^2+\sum_{\bm{k}}\frac{g^2}{2\omega_{\bm{k}}\Omega})x-\sum_{\bm{k}}\frac{g}{\sqrt{2\omega_{\bm{k}}\Omega}}(m_V+m_N+\omega_{\bm{k}})x_{\bm{k}},\label{x eq}\\
    &\ddot{x}_{\bm{k}}=-(m_N+\omega_{\bm{k}})^2x_{\bm{k}}-\frac{g^2}{2\Omega\sqrt{\omega_{\bm{k}}}}\sum_{\bm{k'}}\frac{x_{\bm{k'}}}{\sqrt{\omega_{\bm{k'}}}}-\frac{g}{\sqrt{2\omega_{\bm{k}}\Omega}}(m_V+m_N+\omega_{\bm{k}})x,\label{xk eq}
\end{align}
\end{widetext}
where the summation over $\bm{k}$ and $\bm{k}'$ are cut off at a magnitude of $k_c$. The initial conditions are $x(0)=\dot{x}(0)=\dot{x}_{\bm{k}}(0)=0$, $x_{\bm{k}}(0)=u_{\bm{K}}\delta_{{\bm{k,K}}}$. Eq.~\eqref{x eq} yields the formal solution:
\begin{equation}\label{formal x}
    x=\frac{i}{2\Omega_1}\sum_{\bm{k}}\Omega_{\bm{k}}^2\int_0^te^{i\Omega_1(t-t')}x_{\bm{k}}(t')dt',
\end{equation}
with $\Omega_1^2\equiv m_V^2+\sum_{\bm{k}}g^2/2\omega_{\bm{k}}\Omega$ and $\Omega_{\bm{k}}^2\equiv g(m_V+m_N+\omega_{\bm{k}})/\sqrt{2\omega_{\bm{k}}\Omega}$. Applying integration by parts to $S(\bm{k},t)\equiv\int_0^tx_{\bm{k}}(t')\exp\left[i\Omega_1(t-t')\right]dt'$ in Eq.~\eqref{formal x}, we obtain:
\begin{equation}\label{Skt}
    S(\bm{k},t)=\frac{1}{i\Omega_1}x_{\bm{k}}(t)-\frac{x_{\bm{k}}(0)}{i\Omega_1}e^{i\Omega_1t}-\frac{1}{i\Omega_1}\int_0^t e^{i\Omega_1(t-t')}\dot{x}_{\bm{k}}(t')dt'~.
\end{equation}
 Since $\Omega_1$ diverges as $k_c\rightarrow\infty$, the second term in Eq.~\eqref{Skt} is a rapidly oscillating term that can be neglected in the equation of motion. The third term is also neglectable since it scales as $\Omega_1^{-2}$, which is a higher-order small quantity compared to $x_{\bm{k}}/i\Omega_1$. Retaining only dominant terms gives $S(\bm{k},t)\approx x_{\bm{k}}(t)/i\Omega_1$. Substituting this result into Eq.~\eqref{xk eq} produces the reduced equations of motion for environment oscillators:
\begin{equation}\label{reduce xk}
    \ddot{x}_{\bm{k}}=-(m_N+\omega_{\bm{k}})^2x_{\bm{k}}-\sum_{\bm{k'}}\left(\frac{g^2}{2\Omega\sqrt{\omega_{\bm{k}}\omega_{\bm{k'}}}}+\frac{\Omega_{\bm{k}}^2\Omega_{\bm{k'}}^2}{2\Omega_1^2}\right)x_{\bm{k'}}.
\end{equation}
This reduced equation reveals the cross-excitation between different modes. The coefficients of the last term in Eq.~\eqref{reduce xk} scale as $\Omega^{-1}$, which is small for a large system volume $\Omega$. Thus we have the first order equation by substituting zeroth-order solutions $x_{\bm{k'}}(t)=u_{\bm{K}}\delta_{{\bm{K}},{\bm{k'}}}\exp\left[{i(m_N+\omega_{\bm{k'}})t}\right]$ into Eq.~\eqref{reduce xk} :
\begin{equation}\label{first order xk}
    \ddot{x}_{\bm{k}}=-(m_N+\omega_{\bm{k}})^2x_{\bm{k}}-A_{\bm{k,K}}e^{i(m_N+\omega_{\bm{K}})t}~.
\end{equation}
This equation indicates that each oscillator is driven by a periodic force with frequency $m_N+\omega_{\bm{K}}$ and amplitude $ A_{\bm{k,K}}$~, which is
\begin{equation}\label{amplitude}
    \begin{aligned}
        &A_{\bm{k,K}}=u_{\bm{K}}\left(\frac{g^2}{2\Omega\sqrt{\omega_{\bm{k}}\omega_{\bm{K}}}}+\frac{\Omega_{\bm{k}}^2\Omega_{\bm{K}}^2}{2\Omega_1^2}\right)\\
    &=\frac{u_{\bm{K}}g^2}{2\Omega\sqrt{\omega_{\bm{k}}\omega_{\bm{K}}}}
    \left[1+\frac{(m_V+m_N+\omega_{\bm{k}})(m_V+m_N+\omega_{\bm{K}})}{m_V^2+\sum_{\bm{k'}}g^2/2\omega_{\bm{k'}}\Omega}\right]~.
    \end{aligned}
\end{equation}
Eq.~\eqref{first order xk} implies that only oscillators resonant with the initial $K$-mode oscillator (i.~e.~oscillators with $\omega_{\bm{k}}=\omega_{\bm{K}}$) are significantly excited. The amplitudes of off-resonant modes are neglectable compared to those of the resonance modes after a sufficient long interaction time $t$ . This selective amplification mirrors quantum elastic scattering in the Lee model, where the outgoing waves contain only states with the same energy as that of the in-state.

Furthermore, the cutoff-dependent scattering amplitude in both the classical model and the Lee model also demonstrates parallelism. For the resonant modes ($|\bm{k}|=|\bm{K}|$), the amplitude of driving force $A_{\bm{K,K}}$ decreases as the model cut-off $k_c$ increases. Accordingly, the amplitudes of the resonant environmental oscillators decrease, indicating that the scattering is suppressed as an increasing number of ultraviolet modes are included in the model. This result closely parallels that in the Lee model, which similarly indicates a suppression of scattering amplitudes as the mode cut-off of the bosonic field increases. Especially, in the high frequency limit $|\bm{K}|\to+\infty$ without momentum cutoff, the amplitude $A_{\bm{K,K}}$ vanishes, replicating the Lee model's scattering cross-section disappearance. This confirms that the suppression of scattering is also a classical feature of the model rather than quantum mechanical one.

\section{\label{sec conclusion}Conclusion}
In this paper, we have explored the classical analogue of the Lee model and uncovered several important insights into the foundational structure of mass renormalization. By establishing a correspondence between the Lee model and a classical model of coupled harmonic oscillators with unconventional $xx\text{-}pp$ type interactions, we revealed that the divergence commonly attributed to quantum field theory actually have deep classical roots. A comparison of the Lee model, the CHO model with Hamiltonian~\eqref{config rep} and the generalized TCHO model with Hamiltonian~\eqref{TCHO H} are summarized in Tab.~\ref{tab:results}.
\begin{table*}
    \centering
    \renewcommand{\arraystretch}{1.6} 
    \setlength{\tabcolsep}{0.6em} 
    \begin{tabular}{@{}c|c|c|c@{}} 
        \toprule
        \textbf{Comparison} & \textbf{Lee Model} & \textbf{CHO Model} & \textbf{Generalized TCHO Model} \\ 
        \midrule
        
        Oberservables 
        & Physical mass & Fundamental frequency & Fundamental frequency \\
        [1.5ex] 
        
        Properties & divergent, negative & divergent, negative & finite, negative \\
        [1.5ex]
        
        Renormalization term & $\delta mV^\dagger V$ & $\delta m (x^2 + p^2)/2$ & $\sum_{i=1}^2 (a_i p_i^2 + b_i x_i^2)\!/2$ \\
        [1.5ex]
        
        After renormalization & finite, positive & finite, positive & finite, positive \\
        [1.5ex]
        
        Scattering strength & Suppressed to zero & Suppressed & \dash\\
        \bottomrule
    \end{tabular}

    \caption[Comparison of Lee and CHO models]{A comparison of the Lee model with the CHO model (Hamiltonian~\eqref{config rep}) and the generalized TCHO model (Hamiltonian~\eqref{TCHO H}). This table presents the physical properties of observables as the cutoff frequency (coupling constant in the case of the TCHO model) tends to infinity. The mass renormalization $\delta m$ is given by Eq.~\eqref{d m}. The mass renormalization term $\delta m V^\dagger V$ (Lee model) corresponds to the Fock representation of fundamental frequency renormalization (CHO model). Through renormalization, these observables attain finite positive values. Both the Lee model and the CHO model exhibit scattering strength suppression with increasing cutoff frequency, but with different asymptotic behaviors: the limit in the CHO model scales inversely with the frequency of the initially excited oscillator, whereas that of the Lee model approaches zero. It should be noted that this behavior is absent in the TCHO model because the number of frequencies is fixed as two. The conditions that $a_i$ and $b_i$ must satisfy are given by Eq.~\eqref{p-d condition 1} and Eq.~\eqref{p-d condition 2}.}
    \label{tab:results}
\end{table*}
This reinterpretation not only bridges the conceptual gap between classical and quantum frameworks, but also enriches our understanding of the origins of mass renormalization.
We formulated a generalized renormalization scheme applicable to coupled oscillator systems with phenomenological interaction terms, identifying the mathematical conditions under which such schemes remain consistent. Within this unified framework, the Lee model appears as a special case, thereby situating it within a broader theoretical context.

Moreover, our investigation into the classical counterpart of $N\text{-}\theta$ scattering process revealed a parallel behavior in the scattering strength's dependence on the cutoff frequency that the scattering strength decreases as the cut-off frequency increases. This classical manifestation reflects and supports the quantum field theories based on Lee model, highlighting an unexpected level of structural similarity between quantum scattering systems and their classical analogs.

Looking ahead, we plan to further investigate the theoretical implications of this classical-quantum mapping. Specifically, we will analyze the intersection point in the curve depicted in Fig.~\ref{allowed area} and demonstrate its correspondence to the renormalization scheme of the Dicke model~\cite{dicke1954coherence}. This connection may offer a new perspective on the debated issue of the Dicke phase transition in light-matter interaction systems, potentially providing theoretical support for its absence as suggested in earlier studies~\cite{hepp1973superradiant,wang1973phase,rzazewski1975phase}.

\begin{acknowledgments}
The authors would like to thank Hong Yuan and Ruoxun Zhai for their valuable discussions. This work was supported by the National Natural Science Foundation of China (NSFC) (Grant No. 12088101) and NSAF No. U2330401.
\end{acknowledgments}

\appendix*
\section{The Lee model revisited}
To comprehend the significance of our study, we revisit the Lee model for the mass renormalization~\cite{lee1954some} and the vanishing of the cross-section of the $N$-$\theta$ scattering~\cite{lee1954some,kallen1955mathematical}.

\subsection{Mass renormalization in the Lee model\label{app:mass renormalization}}
 The Hamiltonian of the Lee model is given by Eq.~\eqref{Lee H}. The state of a single $V$ particle is $\ket{V}=V^\dagger\ket{0}$, which belongs to the $N\theta/V$ sector. This sector is spanned by $\{\ket{V},\ket{N\theta_{\bm{k}_1}},\cdots,\ket{N\theta_{\bm{k}_n}}\}$, where $\ket{N\theta_{\bm{k}}}=N^\dagger a_{\bm{k}}^\dagger\ket{0}$. Therefore, the eigenstates of the Lee Hamiltonian can be written as
 \begin{equation}
     \ket{\psi}=Z_2^{1/2}\left[\ket{V}+c_k\ket{N\theta_k}\right],
 \end{equation}
where $Z_2^{1/2}$ is the normalization coefficient. Substituting this expansion into the eigen equation $H_\text{Lee}\ket{\psi}=\omega\ket{\psi}$ yields
\begin{equation}
    \left\{
    \begin{aligned}
        &m_V+\sum_{\bm{k}}\frac{g}{\sqrt{2\omega_{\bm{k}}\Omega}}c_k=\omega~,\\
        &(\omega-m_N-\omega_{\bm{k}})c_k=\frac{g}{\sqrt{2\omega_{\bm{k}}\Omega}}~.
    \end{aligned}
    \right.
\end{equation}
By eliminating $c_k$ we get the following eigenvalue equation
\begin{equation}
     h(\omega) \equiv \omega - m_V - \sum_{\bm{k}} \frac{g^2}{2\Omega\omega_{\bm{k}}(\omega - m_N - \omega_{\bm{k}})} = 0~.
\end{equation}
The function $h(\omega)$ is plotted in Fig.~\ref{Fig secular function}. The lowest solution of this equation, denoted as $m_V'$, corresponds to the physical mass of the $V$ particle, since $m_V'$ approaches $m_V$ as the coupling constant $g$ tends to zero. It is improtant to note that $m_V'<m_N+\mu$, thus in three-dimensional space the summation in $h(\omega)$ diverges for any finite value of $m_V'$. Consequently, $m_V'\to-\infty$, which means that the physical mass of $V$ particle is both negative and divergent. Therefore, the model Hamiltonian $H_{\text{Lee}}$ requires modification, specifically the mass renormalization. By introducing a renormalization term $\delta m V^\dagger V$ into $H_\text{Lee}$ where
\begin{equation}
    \delta m = \sum_{\bm{k}} \frac{g^2}{2\Omega \omega_{\bm{k}} (m_N + \omega_k - m_V)}~,
\end{equation}
the physical mass of the $V$ particle is now positive and finite with value of $m_V$.

\subsection{$N$-$\theta$ scattering\label{app:scattering}}
One of the central goals of QFT lies in determining the $S$-matrix for different scattering processes. Given that scattering cross section correspond to observables, their theoretical calculation enables direct experimental verification of predictions. In the Lee model, the $N\text{-}\theta$ scattering process characterizes the process that a $\theta$ particle with momentum $\bm{k}$ being scattered by an $N$ particle, where the initial quantum state is expressed as $\ket{N\theta_{\bm{k}}} = N^\dagger\theta_{\bm{k}}^\dagger\ket{0}$. The $S$-matrix elements $S_{\bm{k,k'}}$ quantify transition probabilities between in-state $\ket{N\theta_{\bm{k}}}$ and out-states $\ket{N\theta_{\bm{k}'}}$.

Physical scattering states obey the \textit{Lippmann-Schwinger equation}:
\begin{equation}\label{Lippman}
|N\theta_k\rangle^\pm = \lim_{\epsilon\to 0^+} \left(1 + \frac{1}{E_k - H \pm i\epsilon}H_I\right)N^\dagger a_k^\dagger|0\rangle
\end{equation}
where $E_k = m_N + \omega_k$ denotes the asymptotic energy. The $\pm i\epsilon$ prescription establishes appropriate boundary conditions for incoming and outgoing states. Through standard operator algebra manipulations, the in/out states in Eq.~\eqref{Lippman} take the form
\begin{equation}\label{scattering states}
\ket{N\theta_{\bm{k}}}^\pm=\ket{N\theta_{\bm{k}}}+\alpha_{\bm{k}}^\pm\ket{V}+\sum_{\bm{k}'}\beta_{\bm{k,k'}}^\pm\ket{N\theta_{\bm{k'}}}~,
\end{equation}
with coefficients
\begin{align}
    \alpha_{\bm{k}}^\pm&=\frac{g}{\sqrt{2\omega_{\bm{k}}\Omega}}\frac{1}{h(m_N+\omega_{\bm{k}}\pm i\epsilon)}~,\\
    \beta_{\bm{k,k'}}^\pm&=\frac{g}{\sqrt{2\omega_{\bm{k'}}\Omega}}\frac{\alpha_{\bm{k}}}{\omega_{\bm{k}}-\omega_{\bm{k'}}\pm i\epsilon}~.
\end{align}
Here $h(\omega)$ corresponds to the secular function defined in Eq.~\eqref{eigenvalue eq}:
\begin{equation}\label{h diverge}
    \begin{aligned}
    &h(m_N+\omega_{\bm{k}}\pm i\epsilon)\\
    &=m_N+\omega_{\bm{k}} - m_V- \sum_{|\bm{k}'|\le k_c}\frac{g^2}{2\Omega\omega_{\bm{k}'}(\omega_{\bm{k}} -  \omega_{\bm{k}'}+i\epsilon)}.
\end{aligned}
\end{equation}
The final term in Eq.~\eqref{scattering states} describes outgoing scattering components with different momentum $\bm{k'}$. The scattering matrix elements derive from:
\begin{equation}
    \begin{aligned}
        S_{\bm{p,k}} =& \prescript{-}{}{\braket{N\theta_{\bm{p}}}{N\theta_{\bm{k}}}}^+\\
=&\delta^{(3)}(\bm{p-k})-2\pi i\delta(\omega_{\bm{k}}-\omega_{\bm{k}})T_{\bm{p,k}}~,
    \end{aligned}
\end{equation}
where $\rho(\bm{k})=\sum_{\bm{k}'}\delta(\bm{k}-\bm{k'})$ represents the state density near momentum $\bm{k}$, and
\begin{equation}
    T_{\bm{p,k}}=\frac{g^2\rho(\bm{k})}{2\omega_{\bm{k}}\Omega}\frac{1}{h(m_N+\omega_{\bm{k}}+i\epsilon)}~.
\end{equation}
This $T$-matrix directly determines the differential scattering cross-section $\mathrm{d}\sigma/\mathrm{d}\Theta\propto|T_{\bm{p,k}}|^2$, which constitutes an experimentally accessible quantity. Notably, as the momentum cutoff $k_c$ approaches infinity in three-dimensional space, the secular function $h(m_N+\omega_{\bm{k}}\pm i\epsilon)$ develops divergences, leading to suppression and eventual vanishing of the scattering cross-section $\mathrm{d}\sigma/\mathrm{d}\Theta$.

\nocite{*}

\bibliography{Reference}

\end{document}